\documentclass[aps,prl,floatfix,twocolumn,groupedaddress,showpacs,letterpaper,amsmath,amssymb]{revtex4}
\pdfoutput=1
\usepackage{amsfonts,mathrsfs,amsthm}
\usepackage{amsmath,amssymb,graphicx,hyperref}
\hypersetup{pdfnewwindow=true, colorlinks=true, linkcolor=blue, anchorcolor=blue, citecolor=blue, filecolor=blue, menucolor=blue, urlcolor=blue}
\newcommand{\beq}{\begin{equation}}
\newcommand{\eeq}{\end{equation}}
\newcommand{\bea}{\begin{eqnarray}}
\newcommand{\eea}{\end{eqnarray}}
\newcommand{\ds}{\displaystyle}
\newcommand{\nn}{\nonumber}

\newcommand{\1}{{1\hspace*{-0.5ex} \textrm{l} \hspace*{0.5ex}}}

\begin{document}

\title{Hexadecapolar Kondo effect in URu$_2$Si$_2$?}

\author{Anna I.\ T\'oth}\author{Gabriel Kotliar}

\affiliation{Department of Physics \& Astronomy, Rutgers University,
  Piscataway, New Jersey 08854-8019, USA}

\date{\today}

\begin{abstract}
We derive the coupling of a localized hexadecapolar mode to conduction electrons in tetragonal symmetry.
The derivation can be easily adapted to arbitrary multipoles in arbitrary environment.
We relate our model to the two-channel Kondo (2CK) model and show that for an $f^2$-configuration,
a relevant crystal field splitting in addition to the 2CK interaction is intrinsic to tetragonal symmetry.
We discuss possible realizations of a hexadecapolar Kondo effect in URu$_2$Si$_2$. Solving our model
we find good agreement with susceptibility and specific heat measurements in Th$_{1-x}$U$_x$Ru$_2$Si$_2$ ($x\ll1$).
\end{abstract}

\pacs{71.10.Hf, 71.27.+a, 75.20.Hr}

\maketitle

\paragraph{Introduction.}
In an seminal paper \cite{Cox87}, Cox addressed  important differences
between U- and  Ce-based heavy electron materials in terms of the
atomic structure of their  $f$-shell. In Ce-based systems
the most probable configuration has one $f$-electron, in
contrast to many U-based materials having an $f^2$-many body
state as the most probable one. In crystal structures where the
U-site has cubic symmetry, the $f^2$-states can give rise to
quadrupolar degrees of freedom which when coupled to
conduction electrons lead to two-channel Kondo non-Fermi liquid
behavior \cite{Cox87,Nozieres80}.
Meanwhile various multipolar orderings have been observed
\cite{Paixao02} as well as proposed as candidates for ``hidden
order'' (HO) in materials with clear phase transitions but
without an obvious order parameter. A prominent example in this
area is provided by  URu$_2$Si$_2$. For this material
quadrupolar \cite{Santini94}, octupolar \cite{Kiss05},
hexadecapolar \cite{Haule09} and triakontadipolar
\cite{Cricchio09} order parameters have all been put forth.
Recent experiments have definitely ruled out quadrupolar order
\cite{Walker11}, whereas another might implicitly hint at it
\cite{Okazaki11}. On the other hand, the hypothesis of  
active Uranium hexadecapolar degrees of
freedom provides a natural explanation for numerous experiments
\cite{Kusunose11}.  
In this Letter, we generalize the work of ref.\ \cite{Cox87} and present a simple 
construction of low-energy Hamiltonians that describe  
the coupling between multipoles and conduction electrons in the
tetragonal crystal field of URu$_2$Si$_2$. We show that the U  
hexadecapolar degrees of freedom, couple symmetrically to
multiple channels of conduction electrons.  We solve the resulting model 
using the numerical renormalization group, and thereby successfully describe the properties of 
Th$_{1-x}$U$_x$Ru$_2$Si$_2$ ($x\ll1$) \cite{Amitsuka94,Toth10}.
Thus hexadecapolar 
fluctuations also serve as an explanation for the anomalies
observed Th$_{1-x}$U$_x$Ru$_2$Si$_2$ ($x\ll1$).

\paragraph{Construction.}
To construct a tractable model, valid at very low-energies,
we take into account only the two lowest-lying
$f$-configurations with double occupancy. This is  motivated by
recent LDA+DMFT calculations which indicate that, while U has
mixed valence in this material with $f$-electron occupancy
between 2 and 3, two crystal field singlets with double occupancy
and different symmetries have the highest probability
\cite{Haule09}.
The Uranium degrees of freedom are then described by a
$J\,=\,4$ multiplet, split  by the
tetragonal crystal structure.
The ground state and the nearest
excited level are,  respectively, time-reversal and parity even, $\,A_{\,2g}\,$  and $\,A_{\,1g}\,$
basis states of the group $\,{\mathscr{T}}\,\times\,D_{4h}\,$ with
$\,\mathscr{T}\,=\,\{{\cal I},\,{\cal T}\}\,$ the group of time-reversal,
$\,{\cal I}\,$ the identity,
$\,{\cal T}\,$  the time-reversal operator; and $\,D_{4h}\,$ the tetragonal point group including parity. Viz., the lowest-lying singlets are\,:
$\,|A_{\,2g}\rangle\equiv\frac{i}{\sqrt{2}}\left(|4\rangle\,-\,| -4\rangle\right)\,$, and
$\,|A_{\,1g}\rangle\equiv\frac{\cos\phi}{\sqrt{2}}\,\left(| 4\rangle\,+\,| -4\rangle\right)\,+\,\sin\phi\,| 0\rangle\,$,
given in terms of the eigenvectors, $\,|J_z\rangle\,$,  of the operator $\,{\hat J}_z\,$
in the $\,J\,=\,4\,$ multiplet with the quantization axis chosen parallel to the $c$-axis of
the crystal.
To keep the equations short, we follow
refs.\ \cite{Nozieres80,Cox87} and assume, the $\,f$-shell of the \,U\, atom hybridizes predominantly
with $\,l\,=\,3,\,J\,=\,\frac 5 2\,$ conduction electrons.
The conduction electrons at the local site,
$\,\psi^\dagger_{l\,J\,J_z}\,$ can be classified into  the four
double-valued or spinor irreducible representations (irreps),
$\Gamma_{\,6p},\,\Gamma_{\,7p}\,$  of the tetragonal double point
group, $\,{\bar D}_{4h}\,$, with $\,p\,=\,g/u\,$ for parity
even/odd irreps (i.e.\ for $l$ even/odd). Under time-reversal symmetry
$\,{\cal T}\psi^\dagger_{l\,J\,J_z}{\cal T}^{-1}=(-)^{l\,-\,J\,+\,J_z}\,\psi^\dagger_{l\,J\,(-J_z)}\,$.
We set up a basis so that
the $\,\alpha\,=\,\pm\,$ components of the Kramers
doublets, $\,\Psi^\dagger_{\Gamma_{\,jp}^{(n)}\,\alpha}\,$ (where
$n$ enumerates doublets of the same type within the same $J$
multiplet, and $j=6,7$) comply with our convention\,:\,
$\,{\cal T}\Psi^\dagger_{\Gamma_{\,jp}^{(n)}\,+}{\cal T}^{-1}=\Psi^\dagger_{\Gamma_{\,jp}^{(n)}\,-}\,$,
implying the same for annihilation operators.
For the local conduction electron basis,
we choose the following
two independent $\,\Gamma_{\,7u}\,$ Kramers doublets for creation
operators\,:
$\,
\left[
\begin{array}{c}
\Psi^\dagger_{\Gamma_{\,7u}^{(1)}\,+}\\
\Psi^\dagger_{\Gamma_{\,7u}^{(1)}\,-}
\end{array}
\right]\equiv
\left[
\begin{array}{c}
\psi^\dagger_{\frac 5 2}\\
-\,\psi^\dagger_{-\frac 5 2}
\end{array}
\right],\quad
\Psi^\dagger_{\Gamma_{\,7u}^{(2)}}\equiv
\left[
\begin{array}{c}
\psi^\dagger_{-\frac 3 2}\\
-\,\psi^\dagger_{\frac 3 2}
\end{array}
\right],\,$
and one $\,\Gamma_{\,6u}:\,$
$\Psi^\dagger_{\Gamma_{\,6u}^{(1)}}\,\equiv\,\left[
\begin{array}{c}
\psi^\dagger_{\frac 1 2}\\
-\,\psi^\dagger_{-\frac 1 2}
\end{array}
\right]\,,\,$
on using the  condensed notation:
$\psi^\dagger_{J_z}\,\equiv\,\psi^\dagger_{3\,\frac 5 2\,J_z}$.
Adjoint doublets with the same transformation properties in the same basis are
\bea
\Xi^{}_{\Gamma_{\,7u}^{(1)}}\equiv
\left[
\begin{array}{c}
\psi^{}_{-\frac 5 2}\\
\psi^{}_{\frac 5 2}
\end{array}
\right],\,\,
\Xi^{}_{\Gamma_{\,7u}^{(2)}}\equiv
\left[
\begin{array}{c}
\psi^{}_{\frac 3 2}\\
\psi^{}_{-\frac 3 2}
\end{array}
\right],\,\,
\Xi^{}_{\Gamma_{\,6u}^{(1)}}\equiv
\left[
\begin{array}{c}
\psi^{}_{-\frac 1 2}\\
\psi^{}_{\frac 1 2}
\end{array}
\right].
\nn
\eea

Kondo Hamiltonians are made up of spin-flip and diagonal processes:
$\,{\cal H}_{K}\,=\,{\cal H}_\perp\,+\,{\cal H}_z\,$.
When constructing these two parts connecting the two singlets,
the only relevant, non-trivial tensor products of irreps are
$\,\Gamma_{\,6u}\otimes\Gamma_{\,6u}=\Gamma_{\,7u}\otimes\Gamma_{\,7u}=A_{\,1g}\oplus A_{\,2g}\oplus
E_{\,g}\,$ and $\,A_{\,2g}\otimes A_{\,2g}\,=\,A_{\,1g}\,$
\cite{Koster63}.  Inserting the appropriate tetragonal Clebsch--Gordan coefficients,
symmetry thus binds the most general form for
the spin-flip and diagonal parts to be \cite{Koster63}
\bea
{\cal  H}_\perp&=&i\ds{\sum_{n,m=1}^2}{\cal J}_{\perp}^{\,n\,m}
\left(\Psi^\dagger_{\Gamma_{\,7u}^{(n)}\,+}
\Xi^{}_{\Gamma_{\,7u}^{(m)}\,-}+\Psi^\dagger_{\Gamma_{\,7u}^{(n)}\,-}
\Xi^{}_{\Gamma_{\,7u}^{(m)}\,+}\right)\nn\\
&&\times\,|A_{\,1g}\rangle\langle A_{\,2g}|\,
+ \,h.c.,\label{eq:Hperp}\\ \label{eq:Hz}
{\cal  H}_z&=&\ds{\sum_{n,m=1}^2}\,\,\,\ds{\sum_{i\,\in\,\{1,2\}}}\,\,\,{\cal J}_{z}^{\,n\,m\,i}\\
&&\times\left(\Psi^\dagger_{\Gamma_{\,7u}^{(n)}\,+}
\Xi^{}_{\Gamma_{\,7u}^{(m)}\,-}-\Psi^\dagger_{\Gamma_{\,7u}^{(n)}\,-}
\Xi^{}_{\Gamma_{\,7u}^{(m)}\,+}\right)|A_{\,ig}\rangle\langle A_{\,ig}|.\nn
\eea
The couplings are real and must satisfy $\,{\cal J}_{z}^{\,1\,2\,i}\,=\,{\cal J}_{z}^{\,2\,1\,i}\,$ to ensure hermiticity, but otherwise arbitrary.
We omitted processes including $\Gamma_6$ electrons as they decouple from the impurity.
The hexadecapolar, i.e.\ ``spin-flip'' fluctuations are thus coupled to four species of
conduction electrons, namely to the two independent $\Gamma_{7u}$ Kramers doublets.
${\cal H}_\perp$ has the structure of the two-channel 
Kondo (2CK) model where the role of spin index is played by the
index that distinguishes the two different $\Gamma_{7u}$'s, and the channels
are distinguished by the Kramers indices. To make this correspondence more explicit,
we  introduce the operators\,:
$\,\left[\begin{array}{c}\eta_{a\,\uparrow}\\\eta_{a\,\downarrow}\end{array}\right]\equiv\left[\begin{array}{c}\psi_{-\frac3 2}\\
\psi_{\frac 5 2}\end{array}\right]\,,\,
\eta_{\,b}\equiv\left[\begin{array}{c}\psi_{\frac 3 2}\\-\psi_{-\frac 5 2}\end{array}\right]\,$,
and perform the unitary transformation:\,
$\,|A_{\,1g}\rangle\,\rightarrow\,|A_{\,1g}^\prime\rangle\,\equiv\,i\,
|A_{\,1g}\rangle\,$, which allows us to  rewrite ${\cal H}_{\,\perp}$ in the standard notation
\bea\nn
{\cal H}{}_{\,\perp}&=&\frac{{\cal J}_\perp^{\,2\,1}-{\cal J}_\perp^{\,1\,2}}{2}
\,\,\eta^\dagger_{\,q\,\mu}\,\sigma^+_{\mu\,\nu}\,\eta^{}_{\,q\,\nu}\,{S^-}\,
+\,h.c.\,+\,{\cal O}\,{S^x}\,,
\eea
with $q\,\in\,\{a,b\}\,,\,\,\,\mu,\nu\,\in\,\{\uparrow,\downarrow\}\,$.
Here and in the following, repeated channel ($q$) and spin ($\mu,\nu$) indices  are to be summed over;
${\cal O}={\cal O}^\dagger$ contains only conduction
electrons \footnote{ ${\cal O}\,\equiv\,{\cal J}_\perp^{\,12}\,\,\eta^\dagger_{\,q\,\mu}\,\left(\sigma^+_{\mu\,\nu}
\,+\,\sigma^-_{\mu\,\nu}\right)\,\eta^{}_{\,q\,\nu}\\
+\,{\cal J}_\perp^{\,11}\,\left[
\eta^\dagger_{\,a\,\mu}\left(\1_{\mu\,\nu}-\sigma^z_{\mu\,\nu}\right)
\,\eta^{}_{\,a\,\nu}-\eta^\dagger_{\,b\,\mu}\left(\1_{\mu\,\nu}-\sigma^z_{\mu\,\nu}\right)
\,\eta^{}_{\,b\,\nu} \right]\\
+\,{\cal J}_\perp^{\,22}\,\left[
\eta^\dagger_{\,a\,\mu}\left(\1_{\mu\,\nu}+\sigma^z_{\mu\,\nu}\right)
\eta^{}_{\,a\,\nu}-\eta^\dagger_{\,b\,\mu}\left(\1_{\mu\,\nu}+\sigma^z_{\mu\,\nu}\right)
\eta^{}_{\,b\,\nu}
 \right]\,.$};
$\sigma^+\equiv\sigma^x+i\sigma^y$ is composed of Pauli matrices; and
$ S^+\equiv S^x\,+\,iS^y
 \equiv|A_{\,2g}\rangle\langle A_{\,1g}^\prime|;
\,S^{z}\,\equiv\,\left(|A_{\,2g}\rangle\langle
A_{\,2g}|\,-\,|A_{\,1g}^\prime\rangle\langle A_{\,1g}^\prime|\right)/\,2\,;\,
S^-\equiv{S^+}^\dagger\,;$
$\1\equiv|A_{\,2g}\rangle\langle A_{\,2g}|+|A_{\,1g}^\prime\rangle\langle A_{\,1g}^\prime|\,$.
\paragraph{Discussion.}
Channel symmetry is a consequence of time-reversal symmetry.
The operator, $\,{\cal O}\,{S^x}\,$ is irrelevant around the 2CK fixed point
(and marginal in the free fermion scaling regime),
as  shown either by NRG calculations or using conformal field theory
results \cite{Affleck93,long_paper}. It does not destroy the 2CK state, since
it neither breaks  channel symmetry, nor lifts the spin degeneracy.
Thus we must have $\,{\cal J}_{\perp}^{\,1\,2}\,\neq\,{\cal J}_{\perp}^{\,2\,1}\,$ in order
for overscreening to occur.  This asymmetry comes up  naturally e.g.\ if we
start off with a spherical symmetric Anderson
Hamiltonian, perform the Schrieffer--Wolff transformation to arrive
at a Kondo-type of interaction and then project
to the crystal field states, $|A_{\,2g}\rangle,\,|A_{\,1g}\rangle$ at
strong spin-orbit (i.e.\ $jj$) coupling, as described in refs.\ \cite{Nozieres80,Cox87,long_paper}.

The diagonal part, ${\cal H}_z$ cannot lead to non-Fermi liquid behavior by
itself, but it can quite possibly destroy it. Channel symmetry is
preserved by time-reversal symmetry. However, the level degeneracies are lifted by the
crystal field, both between the $\,|4\,A_{\,2/1g}\rangle\,$ states and also in
each screening channel between $\,\Gamma_{7u}^{(n)}\,$ electrons with different
$n$'s.  The dangerous terms are
\bea {\cal  H}^{\,rel}_{zi}&=&\Delta_{imp}\,\,
\eta^\dagger_{\,q\,\mu}\,\eta^{}_{\,q\,\mu}\,\,S_z\,,\label{eq:relzi}\\
{\cal H}^{\,rel}_{zc}&=&\Delta_{cond}\,\,
\eta^\dagger_{\,q\,\mu}\,\sigma_{\mu\nu}^z\,\eta^{}_{\,q\,\nu}\,\,\1\,.\label{eq:relzc}
\eea
Both types of crystal field splittings are relevant around the 2CK fixed
point with scaling dimension $\frac12$ \cite{Affleck93,long_paper}, and present in ${\cal H}_z$ with the
amplitudes $\,\Delta_{imp}\,=\,\left({\cal J}_z^{\,2\,2\,1}+{\cal J}_z^{\,1\,1\,1}-{\cal J}_z^{\,2\,2\,2}-{\cal
  J}_z^{\,1\,1\,2}\right)/\,{2}\,,$ and
$\,\Delta_{cond}\,=\,\left({\cal  J}_z^{\,2\,2\,1}-{\cal J}_z^{\,1\,1\,1}+{\cal J}_z^{\,2\,2\,2}-{\cal J}_z^{\,1\,1\,2}\right)/\,{4}\,.$
In fact, they are the only
possibilities for relevant perturbations, if channel symmetry is intact \cite{Affleck93}.

Thus for this model to exhibit 2CK scaling in some
temperature range,  $\,\Delta_{imp}$ and $ \Delta_{cond}\,$ must fall
below the Kondo scale, $T_K$. This necessarily requires fine-tuning, and the basic assumption of the $A_{\,2g}-A_{\,1g}$
scenario---and, as we show below, of any other doublet-ground state
scenarios---is that this accidental degeneracy
is responsible for the unique behavior of URu$_2$Si$_2$ among the
large number of U-based heavy fermions. LDA+DMFT calculations for URu$_2$Si$_2$
are indeed consistent with this accidental degeneracy on the scale of
$T_K$ \cite{Haule09}.
\begin{figure}
\includegraphics[width=1.\linewidth]{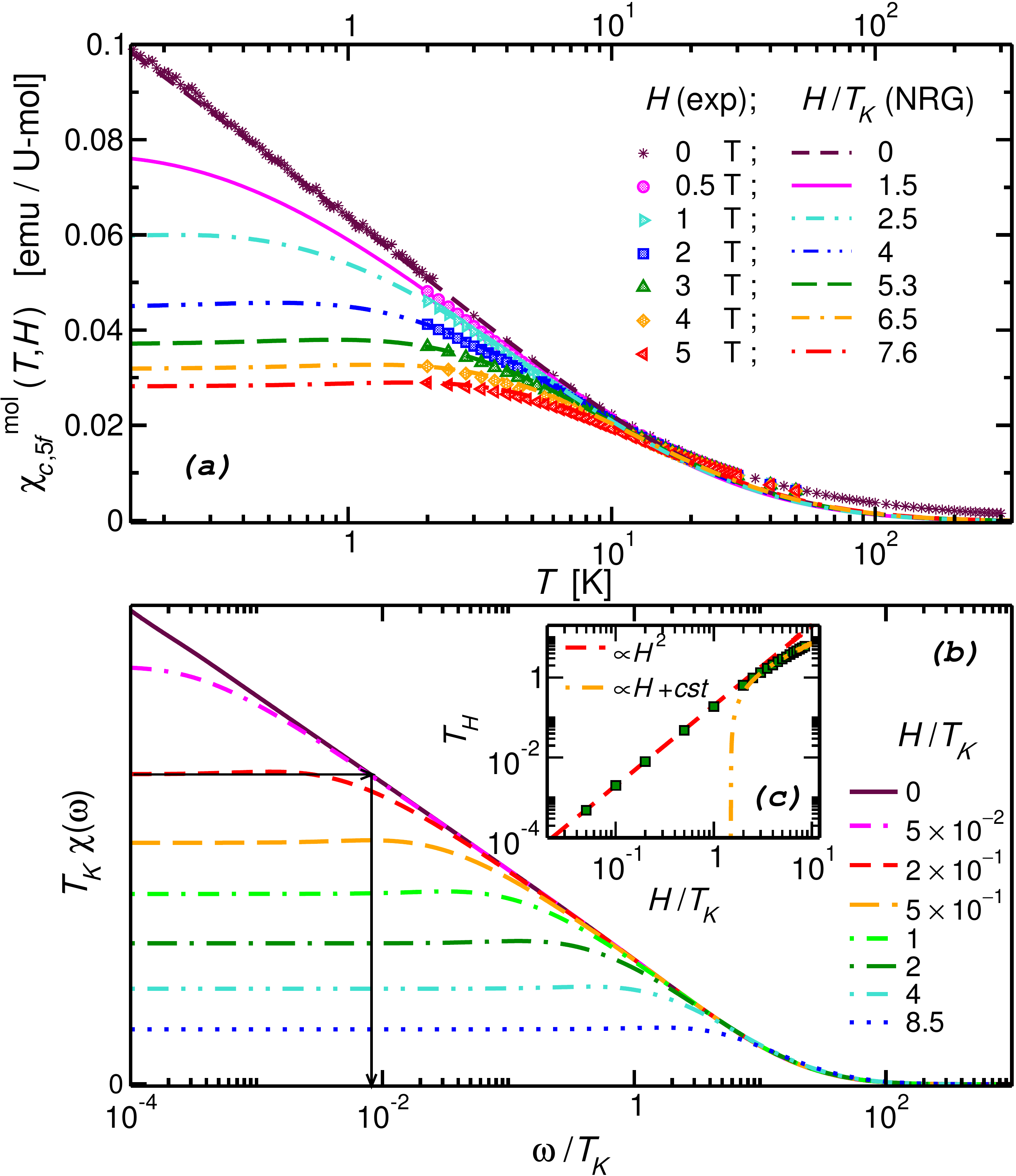}
\caption{$(a):$ Symbols\,: Molar susceptibility, $\chi_{c,5f}^{mol}$, of the
  $5f$ electrons in Th$_{1-x}$U$_x$Ru$_2$Si$_2$
  at $\,x=0.03$ as the function of temperature in magnetic fields between $\,H\,=\,0$
  and 5~T ($H\parallel c$). $(a)-(b):$  Curves\,: The
  local, dynamic susceptibility of the two-channel Kondo model
  in the presence of magnetic field, computed
  with density matrix-NRG at $\,T\,=\,0\,$.  $(c):$ The crossover
  scale, $T_H$ (defined as the intersect of the low-$\omega$ and
  high-$\omega$ asymptotes of $\chi(\omega)$, see plot $(b)$) shows quadratic and linear
  $H$-dependence  in the 2CK and local moment scaling regimes, respectively in good agreement with experiment;
  $T_K$, the crossover scale between the local moment and 2CK scaling
  regimes, is $\approx\,$1.3~K. Notice that the experimental data for $H=0$~T can be equally well fitted
  with every NRG curve where $H\,\lessapprox\,0.2\,T_K$.
}
\label{fig:susc}
\end{figure}

A local, $z$-directed magnetic field results in the following leading
additions to the  Hamiltonian
\bea
{\cal H}_{\,magi}&\propto&\mu_B\,
i\,\left(S^+-S^-\right)\nn\\\nn
{\cal H}_{\,magc}&\propto&\mu_B{\sum_{J_z\in\{\pm\frac32,\pm\frac52\}}}J_z\,\psi^\dagger_{J_z}\psi^{}_{J_z}\1\nn
\eea
for the two-singlet part and for the local conduction electrons, respectively.
These terms have similar effect as ${\cal H}^{rel}_{zi/c}$.
Namely, the impurity part, $\,{\cal H}^{rel}_{zi}\,+\,{\cal H}_{\,magi}\,$, 
amounts to an effective magnetic field (or crystal
field splitting) pointing into other than the $z$-direction. The
same holds true for the conduction electrons with the effective
magnetic field/crystal field splitting being different in the two
channels. Thus while ${\cal H}_K$ is not identical to the 2CK
Hamiltonian,  it flows to the same fixed point when the relevant
perturbations,  which split apart the two different $\Gamma_{7}$
irreps, or the two local singlets, vanish; and the application of
magnetic field thus breaks both the
channel  and the  spin symmetry of the 2CK model.
\paragraph{Comparison with Other Scenarios.}
 For the order of crystal field levels in URu$_2$Si$_2$, other scenarios
have also been put forward in the literature. 
 In turns out that the  structure of the effective 
low-energy Hamiltonian for any two quasi-degenerate states is 
rather similar to that of the two-singlet case considered above.
This applies also to a proposed $E_g$ (or $\Gamma_5$) doublet ground state
\cite{Amitsuka94,Koga95,Ohkawa99,Amitsuka00},
formed by $|E_{g}\,x\rangle,|E_{g}\,y\rangle$.
Fluctuations within this doublet can couple to two products of irreps\,:
to $\Gamma_7\otimes\Gamma_7$ and also to $\Gamma_7\otimes\Gamma_6$.
Importantly, the two different irreps play again the role of spin, while their Kramers indices,
connected by time-reversal, play
again the role of channel index in the 2CK  language.
If the two lower-lying irreps were degenerate, the spin symmetry  of the 2CK model would be unbroken
and 2CK scaling would occur. Since the degeneracy is approximate, the system will eventually flow
to a Fermi liquid fixed point.

In the HO phase of URu$_2$Si$_2$, the  ref.\ \cite{Haule09} proposed order parameter,
$\langle |A_{\,2g}\rangle \langle A_{\,1g}|\rangle$, is non-vanishing due to
its real part which, in highest order of the multipole expansion, contains the expectation value of
the hexadecapolar $A_{2g}$ tensor: $\left[({\hat J}_x^2-{\hat J}_y^2)({\hat
  J}_x{\hat J}_y+{\hat J}_y{\hat J}_x)+({\hat J}_x{\hat J}_y+{\hat J}_y{\hat
 J}_x)({\hat J}_x^2-{\hat J}_y^2)\right]$ \cite{Haule09}.  However, the same reasoning
can be repeated for  $\langle |E_{g}\,x\rangle \langle E_{g}\,y|\rangle$,
whose real part also contains this hexadecapolar ordering.
These points are   substantiated by the construction of the
Hamiltonian for the $E_g$ ground state in the supplementary
material section.
\paragraph{Comparison with Experiment.}
It has long been recognized that  $\chi_c$, the (magnetic)
dipole susceptibility of Th$_{1-x}$U$_x$Ru$_2$Si$_2$ along the
$c$ axis shows $\log T$ behavior at low-$T$ (see refs.\
\cite{Amitsuka94,Amitsuka00} and Fig.\ \ref{fig:susc})
in accord with the 2CK descriptions corresponding to both scenarios. However, susceptibility and
resistivity measurements find that the magnetic field ($H$)
induced crossover scale to a Fermi liquid depends on $H$
linearly, i.e.\ $T_H\propto H^\eta$ with $\eta=1$, which does not
agree with the $\eta=2$ behavior corresponding to the 2CK scaling
regime \cite{Toth10}.
To make  contact with these experiments,
we  solved the model, Eq.\ \eqref{eq:Hperp} by NRG, and confirmed
that it indeed flows to the 2CK fixed point where ${\cal O}S^x$
is irrelevant. Then we added a magnetic field, mimicked
only by Eq.\ \eqref{eq:relzi}, to the 2CK model,  and solved this model
using an upgraded  version of our density matrix-NRG code
detailed in ref.\ \cite{Toth08}.
The values of the magnetic field and the Kondo coupling were adjusted  to  fit  the
experimental data of refs.\ \cite{Amitsuka94,Toth10}.

Invoking $\omega/T$ scaling, we fitted the $T$-dependence of
$\chi_c$ by the dynamic susceptibility of the 2CK model in magnetic field, as we
trust our dynamic correlation functions (produced by the density
matrix algorithm  at $T=0$) better than the thermodynamic
quantities. Fig.\ \ref{fig:susc} shows convincing agreement
between theory and experiment apart from the small discrepancy
for $T>30$~K, i.e.\ for large energies where the resolution of
NRG is limited. We obtained $T_K\approx 1.3$~K from the fit (see
the caption of Fig.\ \ref{fig:susc} for further details on $T_K$).
This finding  places  the measurements in magnetic fields around the crossover region  between the local moment and 2CK
scaling regimes. In both regimes, scale invariance entails the
hyperscaling relation, $\eta\,+\,\nu\,=\,2$ with $\nu$ the
critical exponent defined by $\,\chi\,\propto\,H^{-\nu}$. Thus
for $H=0$, the observed $\nu=0$ gives $\eta=2$, meaning that for
$T$ between 0.1 and 10~K, the system is in the 2CK scaling
regime.  In contrast, for $H=1$ to 5~T,  the experiments measure
$\eta=1$ resulting $\nu=1$. Thus, we conclude, these magnetic
fields in addition to the ubiquitous, relevant crystal field
splitting, are (slightly) larger than $T_K$ and the  system flows
directly from the local moment regime to a one-channel Kondo
fixed point without traversing the 2CK scaling regime.  In Fig.\
\ref{fig:susc}$(a)$, the ratios of magnetic fields to $T_K$,
fitting the susceptibility, further illustrate this point.

\begin{figure}
\includegraphics[width=1.\linewidth]{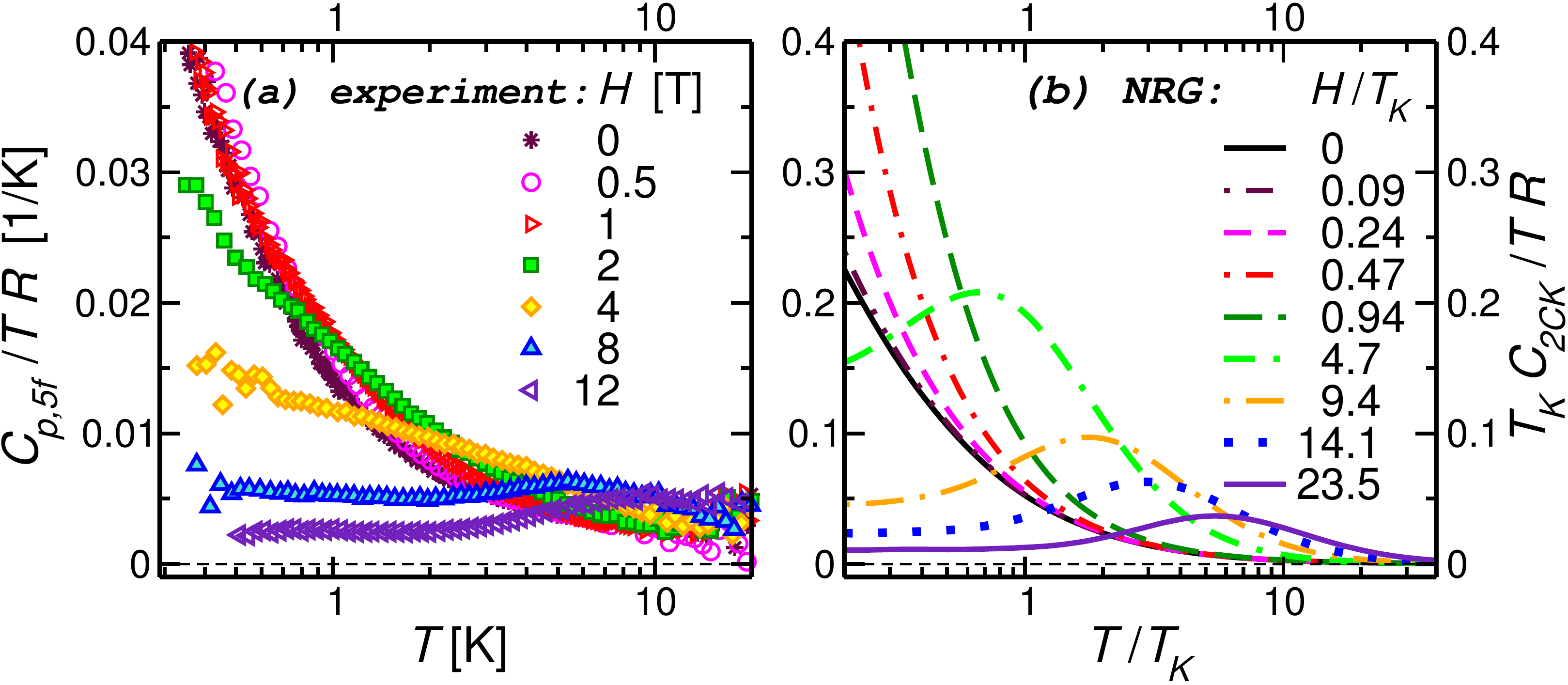}
\caption{$(a):$ The $5f$ electronic specific heat coefficient in units of $R$ ($C_{p,5f}/TR$) of
  Th$_{1-x}$U$_x$Ru$_2$Si$_2$ at $\,x=0.07$, in magnetic fields between 0 and
  12~T, reproduced from ref.\ \cite{Amitsuka00}.
  $(b):$ 2CK specific heat coefficient from NRG for different magnetic field
  values. The two sets of curves display the same trend.
  However there is a factor of $\approx4$ difference in the magnitudes
  (depending on the precise value of $T_K$).  We ascribe this difference to
  experimental inaccuracy, as we observe a factor of $\approx2$ difference between the
  magnitudes of $C_{p,5f}/T$ for the same material published in refs.\ \cite{Amitsuka94} and \cite{Amitsuka00}, respectively.
}
\label{fig:specheat}
\end{figure}

By taking a closer look at the specific heat coefficient
in Fig.\ \ref{fig:specheat}$(a)$, we can reinforce these statements, and
get another estimate for $T_K$. Two regimes for the given magnetic
field values are clearly visible: The curves for $H=0.5$ and 1~T  slightly overshoot the  curve at $H=0$~T
and low-$T$, in contrast to the curves for $H\geq2$~T which exhibit a
bump at around $T\approx H$. The rise of $\gamma\equiv C_{p,5f}/T$ for low-fields at low-$T$ is reminiscent of
the 2CK scaling regime, except that the measured
$T$-dependence of $\gamma$ at low-$T$ for $H=0$ is not quite logarithmic.
These observations can
be explained by the presence of an effective crystal field splitting,
$0<\Delta < T_K$, already at
zero magnetic field, and  placing $T_K$ between 1 and 2 T.
%($\Delta\,+\,$1~T) and
%($\Delta\,+\,$2~T) (with $\Delta\ll T_K$).
From the susceptibility fit we estimate\,: $\Delta$ is anywhere below about $0.2\,T_K$ (c.f.\ Fig.\ \ref{fig:susc}$(b)$).
These assertions are further confirmed by our NRG calculations for the specific heat coefficient
(see Fig.\ \ref{fig:specheat}$(b)$ or Fig.\ 4 in ref.\ \cite{Sacramento90}). For $H\, >\, T_K$, the bumps
at around $T\,\approx\, H$ correspond to the Schottky anomaly due to the
Zeeman splitting between the two local states.

In the 2CK model, if $\,\Delta\,<\,T_K\,$, there
is a non-Fermi liquid (NFL) region over $\,2\,\log\left(\frac{T_K}{\Delta}\right)\,$
decades, since the splitting induced crossover scale to a Fermi liquid
depends on $\,\Delta\,$ quadratically. Susceptibility measurements find an NFL
region over at least one decade, putting an upper bound on the ratio,
$\Delta/T_K < 0.6$ and giving the conservative estimate: $T_K < 5$~T and $\Delta<3$~T.
\paragraph{Conclusion.}
Motivated by recent findings on the electronic structure of URu$_2$Si$_2$ \cite{Haule09},
we derived the Kondo coupling of
localized hexadecapolar fluctuations to conduction electrons
in tetragonal crystal field. The derivation can easily be adapted to arbitrary situations.
The coupling has the form of the 2CK model  plus relevant, spin
symmetry breaking perturbations present even in magnetic zero field.
The local degrees of freedom are symmetrically coupled to two
different irreps of conduction electrons.
Solving the model with NRG we showed that
the hypothesis introduced in ref.\ \cite{Haule09} to describe the hidden order in URu$_2$Si$_2$
can consistently account for the behavior of URu$_2$Si$_2$ in the dilute limit,
Th$_{1-x}$U$_x$Ru$_2$Si$_2$ ($x\ll1$). Nonetheless this behavior, while confirms that the material is strongly correlated,
does not discriminate between different competing scenarios
for the ground state--excited state sequence in this material.
Since the two irreps of conduction electrons are not connected by symmetry, we expect a Fermi liquid to emerge at sufficiently
low-energies. Hence the intermediate non-Fermi liquid regime,
observed in URu$_2$Si$_2$, is a result of an accidental degeneracy.
This accidental degeneracy is responsible for the unique properties of this compound in the dilute and dense limits,
among the hundreds of known U-based heavy fermion materials.
We found that the scale of the crystal field splitting and the Kondo temperature is smaller
in Th$_{1-x}$U$_x$Ru$_2$Si$_2$ ($x\ll1$) than in  URu$_2$Si$_2$.
The splitting between the two $\Gamma_7$ irreps should be sensitive to the conduction
electron filling, and we expect it to be larger in
La$_{1-x}$U$_x$Ru$_2$Si$_2$ \cite{Marumoto96} where clear Fermi liquid behavior is observed.
In this context it is also worth pointing out that in
Th$_{1-x}$U$_x$Ru$_2$Si$_2$ ($x\ll1$), the resistivity
follows an approximate  $\log T$ behavior with a negative
coefficient suggesting that a crossover to a Fermi liquid
behavior takes place at sufficiently low temperatures. The
study of the resistivity, however, will likely require a more
realistic model for the diluted URu$_2$Si$_2$, including all bands
present in the solid; also a more sophisticated approach to calculating the
resistivity in non-Fermi liquid quantum impurity models than the ones present in the literature
\cite{Costi00}; and corrections to the very dilute  limit as was done recently in the context of the NMR experiments \cite{Pezzoli11}.

\paragraph{Acknowledgment.} We thank Piers Coleman for directing us
to refs.\ \cite{Amitsuka94,Koga95,Ohkawa99,Amitsuka00}. We are grateful to him and to Premala Chandra,
Kristjan Haule and Hiroshi Amitsuka for many useful discussions. This
research has been supported by the NSF grant DMR-0906943.

\onecolumngrid

\pagebreak

\section{Supplementary Material for ``Hexadecapolar Kondo effect in URu$_2$Si$_2$?'' }

\subsection{Kondo Hamiltonian for a Tetragonal $E_g$ (or $\Gamma_5$) Doublet Ground State}
The structure of the Hamiltonian for two singlets, $|A_{\,2g}\rangle,|A_{\,1g}\rangle$
and for any other two quasi-degenerate states has many common features.
The proposed $E_g$ (or $\Gamma_5$) doublet for the U ion ground state in Th$_{1-x}$U$_x$Ru$_2$Si$_2$ ($x\ll1$)
is claimed to provide  full symmetry protection for the 2CK state \cite{Amitsuka94,Koga95,Ohkawa99,Amitsuka00}.
We show that crystal field fine-tuning is  needed in this case as well, to hit the 2CK scaling regime.
On the vector space spanned by the $E_g$ ground state doublet
\bea\label{eq:Gamma5_grst_1}\nn
|E_{g}\,x\rangle&\equiv&\frac{1}{\sqrt{2}}\left[\cos\theta\left(|-1\rangle-|1\rangle\right)+\sin\theta\left(|-3\rangle-|3\rangle\right)\right],\\
\label{eq:Gamma5_grst_2}\nn
|E_{g}\,y\rangle&\equiv&\frac{i}{\sqrt{2}}\left[\cos\theta\left(|-1\rangle+|1\rangle\right)-\sin\theta\left(|-3\rangle+|3\rangle\right)\right],
\eea
we have four independent operators defined e.g.\ as
$\1_E\,\equiv\,|E_{g}\,x\rangle\,\langle
E_{g}\,x|\,+\,|E_{g}\,y\rangle\,\langle E_{g}\,y|\,,$
$\sigma^y_E\,\equiv\,i\,\left(|E_{g}\,x\rangle\,\langle
E_{g}\,y|\,-\,|E_{g}\,y\rangle\,\langle E_{g}\,x|\right)\,,$
$\sigma^z_E\,\equiv\,|E_{g}\,x\rangle\,\langle
E_{g}\,x|\,-\,|E_{g}\,y\rangle\,\langle E_{g}\,y|\,$ and
$\sigma^x_E\,\equiv\,|E_{g}\,x\rangle\,\langle
E_{g}\,y|\,+\,|E_{g}\,y\rangle\,\langle E_{g}\,x|\,$. They
transform as $A_{\,1g},\,A_{\,2g},\,B_{\,1g}\,$ and
$\,B_{\,2g}\,$ tensors of $\,D_{4h}\,$, respectively
\cite{Koster63}. In the Kondo limit, these operators can appear
in the Hamiltonian only in certain combinations with the
conduction electrons \cite{long_paper}. Namely, the $\,B_{\,1g}\,$
and $\,B_{\,2g}\,$ tensors can only couple to products of
$\,\Gamma_{6}^{}\,$ and $\,\Gamma_{7}^{}\,$ electrons following
the rule: $\,\Gamma_{\,6u}\otimes\Gamma_{\,7u}\,=\,B_{\,1g}\oplus
B_{\,2g}\oplus E_{\,g}\,$, i.e.\
\bea\label{eq:Hperp_Gamma5}\nn
{\cal H}_\perp^{E_g}&=&i\,\ds{\sum_{m=1}^2}\,{\cal J}^{m}_{\perp B_2}\,
\left(\Psi^\dagger_{\Gamma_{6}^{(1)}+}\,
\Xi^{}_{\Gamma_{7}^{(m)}-}\,+\,\Psi^\dagger_{\Gamma_{6}^{(1)}-}\,
\Xi^{}_{\Gamma_{7}^{(m)}+}\right)\,\sigma^x_{E}\,
\,+\,\ds{\sum_{n,m=1}^2}\,{\cal J}^{\,n\,m}_{\perp\,A_2}\,
\left(\Psi^\dagger_{\Gamma_{\,7}^{(n)}\,+}\,
\Xi^{}_{\Gamma_{7}^{(m)}-}\,+\,\Psi^\dagger_{\Gamma_{7}^{(n)}-}\,
\Xi^{}_{\Gamma_{7}^{(m)}+}\right)\,\sigma^y_{E}\nn\\\nn
&&+\,{\cal J}^{}_{\perp\,A_2}
\left(\Psi^\dagger_{\Gamma_{\,6}^{(1)}\,+}\,
\Xi^{}_{\Gamma_{6}^{(1)}-}\,+\,\Psi^\dagger_{\Gamma_{6}^{(1)}-}\,
\Xi^{}_{\Gamma_{6}^{(1)}+}\right)\,\sigma^y_{E}\,+\,h.c.,
\\
{\cal H}_z^{E_g}&=&
\ds{\sum_{m=1}^2}\,{\cal J}^{m}_{z B_1}\,
\left(\Psi^\dagger_{\Gamma_{6}^{(1)}+}\,
\Xi^{}_{\Gamma_{7}^{(m)}-}\,-\,\Psi^\dagger_{\Gamma_{6}^{(1)}-}
\,\Xi^{}_{\Gamma_{7}^{(m)}+}\right)\,\sigma^z_{E}\,
+\,\ds{\sum_{n,m=1}^2}\,{\cal J}^{\,n\,m}_{z\,A_1}\,
\left(\Psi^\dagger_{\Gamma_{\,7}^{(n)}\,+}\,
\Xi^{}_{\Gamma_{7}^{(m)}-}\,-\,\Psi^\dagger_{\Gamma_{7}^{(n)}\,-}\,
\Xi^{}_{\Gamma_{7}^{(m)}+}\right)\,\1_{E}\nn\\
&&+\,{\cal J}^{}_{z\,A_1}\,
\left(\Psi^\dagger_{\Gamma_{\,6}^{(1)}\,+}\,
\Xi^{}_{\Gamma_{6}^{(1)}-}\,-\,\Psi^\dagger_{\Gamma_{6}^{(1)}\,-}\,
\Xi^{}_{\Gamma_{6}^{(1)}+}\right)\,\1_{E}\,+\,h.c.\,,\nn
\eea
with arbitrary real couplings.
Again, each allowed term describes a
2CK screening process, and  channel symmetry is guaranteed by
time-reversal symmetry. One apparent difference from the
two-singlet case is that  $\Gamma_{6u}$ conduction electrons combined with $\Gamma_{7u}$'s can
also participate in real 2CK screening processes. Besides there remains also the possibility of the
two $\Gamma_{7u}$'s forming the two screening channels.
Once we single out the two lowest-lying channels, the two cases become
very similar in  that the  diagonal processes for the
$E_g$ ground state also include a relevant splitting between the
different crystal field channels of conduction electrons.  Thus
an $E_g$ doublet ground state does not enjoy level degeneracy
protection, and  in this respect, it is not distinguished from
other, two-singlet  ground state scenarios in tetragonal symmetry.
Full protection of the 2CK state can be achieved in a setting were cubic
symmetry merges the $\Gamma_{6} $ and $ \Gamma_7$ representation
into a $\Gamma_8$ quartet as in Cox's original proposal \cite{Cox87}.


\vspace{-2.2em}

\begin{thebibliography}{9}
\bibitem{Cox87} D.\ L.\ Cox, Phys.\ Rev.\ Lett.\ {\bf 59}, 1240 (1987).
\bibitem{Nozieres80} Ph.\ Nozi\`eres, A.\ Blandin, J.\ Phys.\ Paris {\bf 41}, 193 (1980).
\bibitem{Paixao02} For example J. A. Paix{\~ a}o {\it et al.}, Phys.\ Rev.\ Lett.\ {\bf 89}, 187202 (2002);
D. Mannix  {\it et al.}, ibid.\ {\bf 95}, 117206 (2005).
\bibitem{Santini94} P.\ Santini, G. Amoretti, Phys.\ Rev.\ Lett.\ {\bf 73} 1027 (1994).
\bibitem{Kiss05} A.\ Kiss, P.\ Fazekas, Phys.\ Rev.\ B {\bf 71}, 054415 (2005).
\bibitem{Haule09} K.\ Haule, G.\ Kotliar, Nature Physics {\bf 5}, 796 (2009).
\bibitem{Cricchio09} F.\ Cricchio, F.\ Bultmark, O.\ Gr\r{a}n\"as, L.\ Nordstr\"om,
Phys.\ Rev.\ Lett.\ {\bf 103}, 107202 (2009).
\bibitem{Walker11} H.\ C.\ Walker, R.\ Caciuffo, D.\ Aoki, F.\ Bourdarot,
G.\ H.\ Lander, J.\ Flouquet, Phys.\ Rev.\ B {\bf 83}, 193102 (2011).
\bibitem{Okazaki11} R.\ Okazaki {\it et al.}, Science {\bf 331}, 439 (2011).
\bibitem{Kusunose11} H.\ Kusunose, H.\ Harima, \href{http://arxiv.org/abs/1104.2374}{arXiv:1104.2374} (unpublished).
\bibitem{Haule10} K.\ Haule, G.\ Kotliar, EPL {\bf 89}, 57006 (2010).
\bibitem{Amitsuka94} H. Amitsuka, T. Sakakibara, J. Phys. Soc. Jpn. {\bf 63}, 736 (1994).
\bibitem{Toth10} A.\ T\'oth, P.\ Chandra, P.\ Coleman, G.\ Kotliar,
  H.\ Amitsuka, Phys.\ Rev.\ B {\bf 82}, 235116 (2010).
\bibitem{long_paper} A.\ I.\ T\'oth, P.\ Coleman, G.\ Kotliar, unpublished.
\bibitem{Koster63} G.\ F.\ Koster {\it et al.},
{\it Properties of the Thirty-Two Point Groups}, MIT Press, Cambridge, Massachusetts (1963).
\bibitem{Affleck93} I.\ Affleck, A.\ W.\ W.\ Ludwig,  Phys.\ Rev.\ B {\bf 48}, 7297 (1993).
\bibitem{Koga95} M.\ Koga, H.\ Shiba, J. Phys. Soc. Jpn. {\bf 64}, 4345 (1995).
\bibitem{Ohkawa99} F.\ J.\ Ohkawa, H.\ Shimizu, J. Phys.: Cond. Mat. {\bf 11},  L519-L524 (1999).
\bibitem{Amitsuka00} H. Amitsuka {\it et al.}, Physica B {\bf 281}, 326-331 (2000).
\bibitem{Toth08} A.\ I.\ T\'oth, C.\ P.\ Moca, \"O.\ Legeza, and G.\ Zar\'and, Phys.\ Rev.\ B {\bf 78}, 245109 (2008).
\bibitem{Sacramento90} P.\ D.\ Sacramento, P.\ Schlottman, Physica B {\bf  163}, 231-233 (1990).
\bibitem{Marumoto96} K. Marumoto, T. Takeuchi, Y. Miyako, Phys. Rev. B {\bf 54}, 12194 (1996).
\bibitem{Costi00} T. A. Costi, Phys. Rev. Lett. {\bf 85}, 1504 (2000);
  L.\ Borda, L.\ Fritz, N.\ Andrei, G.\ Zar\'and, Phys. Rev. B {\bf 75}, 235112 (2007).
\bibitem{Pezzoli11} M. E. Pezzoli, M. J. Graf, K. Haule, G. Kotliar, A. V. Balatsky,
Phys. Rev. B {\bf 83}, 235106 (2011).
\end{thebibliography}
\end{document}